\documentstyle[preprint,tighten,aps,psfig]{revtex}
\begin{document}

\preprint{FERMILAB Pub-96/115-E,  D\O\ Note 2960,  June 1996}

\title{Search for Anomalous $WW$ and $WZ$ Production in $p\bar p$ Collisions
       at $\sqrt{s}=1.8$ TeV}

\vspace{-0.4in}

%
%
\author{                                                                        
S.~Abachi,$^{14}$                                                               
B.~Abbott,$^{28}$                                                               
M.~Abolins,$^{25}$                                                              
B.S.~Acharya,$^{43}$                                                            
I.~Adam,$^{12}$                                                                 
D.L.~Adams,$^{37}$                                                              
M.~Adams,$^{17}$                                                                
S.~Ahn,$^{14}$                                                                  
H.~Aihara,$^{22}$                                                               
J.~Alitti,$^{40}$
% ...
\\                                                                              
\vskip 0.50cm                                                                   
\centerline{(D\O\ Collaboration)}                                               
\vskip 0.50cm                                                                   
}                                                                               
\address{                                                                       
\centerline{$^{1}$Universidad de los Andes, Bogot\'{a}, Colombia}               
\centerline{$^{2}$University of Arizona, Tucson, Arizona 85721}                 
\centerline{$^{3}$Boston University, Boston, Massachusetts 02215}               
\centerline{$^{4}$Brookhaven National Laboratory, Upton, New York 11973}        
\centerline{$^{5}$Brown University, Providence, Rhode Island 02912}             
\centerline{$^{6}$Universidad de Buenos Aires, Buenos Aires, Argentina}         
\centerline{$^{7}$University of California, Davis, California 95616}            
\centerline{$^{8}$University of California, Irvine, California 92717}           
\centerline{$^{9}$University of California, Riverside, California 92521}        
\centerline{$^{10}$LAFEX, Centro Brasileiro de Pesquisas F{\'\i}sicas,          
                  Rio de Janeiro, Brazil}                                       
\centerline{$^{11}$CINVESTAV, Mexico City, Mexico}                              
\centerline{$^{12}$Columbia University, New York, New York 10027}               
\centerline{$^{13}$Delhi University, Delhi, India 110007}                       
\centerline{$^{14}$Fermi National Accelerator Laboratory, Batavia,              
                   Illinois 60510}                                              
\centerline{$^{15}$Florida State University, Tallahassee, Florida 32306}        
\centerline{$^{16}$University of Hawaii, Honolulu, Hawaii 96822}                
\centerline{$^{17}$University of Illinois at Chicago, Chicago, Illinois 60607}  
\centerline{$^{18}$Indiana University, Bloomington, Indiana 47405}              
\centerline{$^{19}$Iowa State University, Ames, Iowa 50011}                     
\centerline{$^{20}$Korea University, Seoul, Korea}                              
\centerline{$^{21}$Kyungsung University, Pusan, Korea}                          
\centerline{$^{22}$Lawrence Berkeley National Laboratory and University of      
                   California, Berkeley, California 94720}                      
\centerline{$^{23}$University of Maryland, College Park, Maryland 20742}        
\centerline{$^{24}$University of Michigan, Ann Arbor, Michigan 48109}           
\centerline{$^{25}$Michigan State University, East Lansing, Michigan 48824}     
\centerline{$^{26}$Moscow State University, Moscow, Russia}                     
\centerline{$^{27}$University of Nebraska, Lincoln, Nebraska 68588}             
\centerline{$^{28}$New York University, New York, New York 10003}               
\centerline{$^{29}$Northeastern University, Boston, Massachusetts 02115}        
\centerline{$^{30}$Northern Illinois University, DeKalb, Illinois 60115}        
\centerline{$^{31}$Northwestern University, Evanston, Illinois 60208}           
\centerline{$^{32}$University of Notre Dame, Notre Dame, Indiana 46556}         
\centerline{$^{33}$University of Oklahoma, Norman, Oklahoma 73019}              
\centerline{$^{34}$University of Panjab, Chandigarh 16-00-14, India}            
\centerline{$^{35}$Institute for High Energy Physics, 142-284 Protvino, Russia} 
\centerline{$^{36}$Purdue University, West Lafayette, Indiana 47907}            
\centerline{$^{37}$Rice University, Houston, Texas 77251}                       
\centerline{$^{38}$Universidade Estadual do Rio de Janeiro, Brazil}             
\centerline{$^{39}$University of Rochester, Rochester, New York 14627}          
\centerline{$^{40}$CEA, DAPNIA/Service de Physique des Particules, CE-SACLAY,   
                   France}                                                      
\centerline{$^{41}$Seoul National University, Seoul, Korea}                     
\centerline{$^{42}$State University of New York, Stony Brook, New York 11794}   
\centerline{$^{43}$Tata Institute of Fundamental Research,                      
                   Colaba, Bombay 400005, India}                                
\centerline{$^{44}$University of Texas, Arlington, Texas 76019}                 
\centerline{$^{45}$Texas A\&M University, College Station, Texas 77843}         
}                                                                               


\maketitle

\vspace{-0.2in}

\normalsize
\begin{abstract}

\vspace{-0.2in}

We present results from a search for anomalous $WW$ and $WZ$ production in
$p\bar{p}$ collisions at $\sqrt{s}=1.8$~TeV. We used $p\bar{p}\to e\nu jj X$
events observed during the 1992--1993 run of the Fermilab Tevatron collider,
corresponding to an integrated luminosity of $13.7\pm 0.7\ {\rm pb}^{-1}$. A
fit to the transverse momentum spectrum of the $W$ boson yields direct limits
on the {\it{CP}}-conserving anomalous $WW\gamma$ and $WWZ$ coupling parameters
of $-0.9<\Delta\kappa<1.1$ (with $\lambda=0$) and $-0.6<\lambda<0.7$ (with
$\Delta\kappa=0$) at the 95\% confidence level, for a form factor scale
$\Lambda=1.5$~TeV, assuming that the $WW\gamma$ and $WWZ$ coupling parameters
are equal.


\end{abstract}

\input psfig


\newpage

Self-interactions of the electroweak gauge bosons such as $WW\gamma$ and $WWZ$
are a direct consequence of the non-Abelian gauge symmetry of the Standard
Model (SM). The strength of these interactions can be directly measured by
studying gauge boson pair production in $p\bar{p}$ collisions~\cite{Hagiwara}.
For example, $WW$ production is sensitive to the $WW\gamma$ and $WWZ$
couplings, and $WZ$ production is sensitive to the $WWZ$ coupling. Any
deviation of these couplings from their SM values indicates physics beyond the
SM.

The $WW\gamma$ and $WWZ$ interactions are generally described by a Lagrangian
with fourteen independent coupling parameters~\cite{Peccei}, of which
$\kappa_\gamma$ and $\lambda_\gamma$ for the $WW\gamma$ vertex and $\kappa_Z$
and $\lambda_Z$ for the $WWZ$ vertex are usually studied. These are
{\it{CP}}-conserving coupling parameters. In the SM,
$\Delta\kappa_\gamma(\equiv \kappa_\gamma
-1)=\lambda_\gamma=\Delta\kappa_Z(\equiv \kappa_Z -1)= \lambda_Z=0$, and the
production cross section for $p\bar{p}\to W^+W^-X ~(W^\pm ZX)$ at
$\sqrt{s}=1.8$~TeV is 9.5 (2.5)~pb~\cite{Hagiwara}.

Non-SM (i.e.,~anomalous) couplings dramatically increase the production cross
section, and enhance the transverse momentum spectrum of the $W$ boson
($p_T^W\!$) for large values of $p_T^W\!$. Therefore, a study of the $p_T^W\!$
spectrum of $WW(WZ)$ production leads to a sensitive test of the size of the
$WW\gamma$ and $WWZ$ couplings. With anomalous couplings, some helicity
amplitudes of the $p\bar{p}\to WW(WZ)$ processes grow with $\hat{s}$, the
square of the invariant mass of the $WW(WZ)$ system, and cause the cross
section eventually to violate tree level S--matrix unitarity. To avoid this,
the anomalous couplings are commonly parametrized as dipole form factors with a
cut-off scale $\Lambda$:
$\Delta\kappa(\hat{s})={{\Delta\kappa}/{(1+\hat{s}/\Lambda^2)^2}},
~\lambda(\hat{s})={{\lambda}/{(1+\hat{s}/\Lambda^2)^2}}$. Consequently, limits
on the couplings $\Delta\kappa$ and $\lambda$ are dependent on the choice of
$\Lambda$.

The D\O\ collaboration has reported limits on $WW\gamma$ and $WWZ$ anomalous
couplings from two processes using data from the 1992--1993 Fermilab Tevatron
collider run with $p\bar{p}$ collisions at $\sqrt{s}=1.8$~TeV: on the
$WW\gamma$ coupling from a measurement of $W\gamma$ production~\cite{D0:Wgamma}
and on the $WW\gamma$ and $WWZ$ couplings from a search for $W$ boson pair
production in the dilepton decay modes~\cite{D0:WW}. In this letter we present
a new, independent, determination of limits on the $WW\gamma$ and $WWZ$
anomalous couplings obtained from a search for $p\bar{p}\to WWX$ followed by
$W\to e\nu$ and $W\to jj$, and $p\bar{p}\to WZX$ followed by $W\to e\nu$ and
$Z\to jj$, where $j$ represents a jet. Owing to the limited jet energy
resolution, we cannot distinguish $WZ$ events from $WW$ events. This analysis
uses the same data set as in the papers cited above, corresponding to an
integrated luminosity of $13.7\pm 0.7\ {\rm pb}^{-1}$. The CDF collaboration
has reported a similar measurement~\cite{CDF:WW}.

The D\O\ detector and data collection systems are described in
Ref.~\cite{d0:nim}. The basic elements of the trigger and reconstruction
algorithms for jets, electrons and neutrinos are given in Ref.~\cite{TOP_PRD}.

The $WW,WZ\to e\nu jj$ candidates were selected by searching for events
containing a $W\to e\nu$ decay and at least two jets consistent with $W\to jj$
or $Z\to jj$. The data sample was obtained with a trigger which required an
isolated electromagnetic (EM) calorimeter cluster with transverse energy
$E_T>20$~GeV. In offline event selection, this EM cluster was required to be
within $|\eta|\leq 1.1$ in the central calorimeter, or $1.5\leq |\eta|\leq 2.5$
in the end calorimeters, where $\eta$ is the pseudorapidity, defined as
$\eta=-\ln[\tan\theta/2]$ and $\theta$ is the polar angle with respect to the
proton beam direction. Such an EM cluster was identified as an electron if
(i)~the ratio of EM energy to the total shower energy was greater than 0.9;
(ii)~the lateral and longitudinal shower shapes were consistent with those of
an electron; (iii)~the isolation variable of the cluster was less than 0.1,
where isolation is defined as $I=(E_{\rm tot}(0.4)-E_{\rm EM}(0.2))/E_{\rm EM}
(0.2)$, and $E_{\rm tot}(0.4)$ is the total calorimeter energy inside a cone of
radius ${\cal R}\equiv\sqrt{(\Delta\eta)^2+(\Delta\phi)^2}=0.4$, where $\phi$
is the azimuthal angle around the beam axis and $E_{\rm EM}(0.2)$ is the EM
energy inside a cone of radius 0.2; and (iv)~a matching track was found in the
drift chambers. The $W\to e\nu$ decay was identified by an electron with
$E_T^e>25$~GeV and missing transverse energy $\rlap{\,/}E_T>25$~GeV forming a 
transverse mass $M_T^{e\nu}=[2E_T^e \rlap{\,/}E_T (1-\cos\phi^{e\nu})/c^4]
^{1/2}>40$~GeV$/c^2$, where $\phi^{e\nu}$ is the angle between the
$\stackrel{\rightarrow} {E_T^e}$ and $\rlap{\,/}
{\stackrel{\rightarrow}E_T}$~vectors.

Jets were reconstructed using a cone algorithm with radius ${\cal R}=0.3$, and
were required to be within $|\eta|<2.5$. The jet energies were corrected for
detector effects: jet energy scale calibration and out-of-cone showering; for
energy from the underlying event; and for energy loss due to out-of-cone gluon
radiation~\cite{TOP_PRD}. We required that a candidate event contain at least
two jets with $E_T^j>20$~GeV and that the dijet invariant mass (the largest
invariant mass if there were more than two jets with $E_T^j>20$~GeV in the
event) satisfy 50~GeV$/c^2<m_{jj}<110$~GeV$/c^2$, consistent with $W$ and $Z$
boson masses. Monte Carlo studies showed that the standard deviation of the
dijet invariant mass distribution of the signal events is 15~GeV$/c^2$. The
above selection criteria yielded 84 candidate events.

\begin{figure}[htbp]
\centerline{\psfig{file=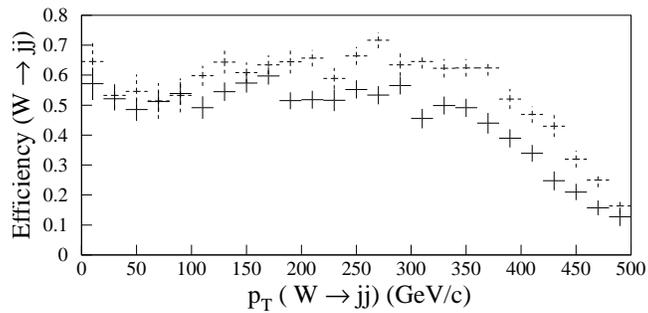
,bbllx=0pt,bblly=390pt,bburx=550pt,bbury=660pt,width=9cm}}
\caption{Total efficiency for $W\to jj$ selection as a function of $p_T^W\!$,
estimated using the {\sc isajet} (solid) and {\sc pythia} (dashed) generators
followed by a full detector simulation.}
\label{JET_EFF}
\end{figure}

The trigger and electron selection efficiencies were measured~\cite{Xsection}
using $Z\to ee$ events. The product of these efficiencies was found to be
$0.78\pm0.02$ in the central calorimeter and $0.62\pm0.01$ in the end
calorimeters. The $W\to jj$ selection efficiency was parametrized as a function
of $p_T^W\!$, as shown in Fig.~\ref{JET_EFF}. This was estimated using events
generated with the {\sc isajet}~\cite{isajet} and {\sc pythia}~\cite{pythia}
programs, followed by a detailed simulation of the D\O\ detector based on the
{\sc geant} package~\cite{geant} and application of our event selection
criteria. The roll-off of efficiency in the $p_T^W\!$ region beyond
$\sim350$~GeV$/c$ is due to merging of the two jets from the $W$ or $Z$ boson.
The use of a cone size as narrow as ${\cal R}=0.3$ for jet reconstruction
ensures that the loss of efficiency occurs only for $W$ and $Z$ bosons with
transverse momenta greater than expected from the coupling parameter values
studied. The $Z\to jj$ efficiency was obtained in a similar manner. In
estimating the detection efficiencies of the $WW(WZ)$ process, we used the
$W(Z)\to jj$ efficiency obtained from {\sc isajet}, which is smaller than that
from {\sc pythia} and therefore gives more conservative limits on the coupling
parameters.


We calculated the overall event selection efficiency as a function of the
coupling parameters using a fast detector simulation program which incorporates
the efficiencies described above and the detector resolutions~\cite{TOP_PRD}.
The $WW(WZ)$ events were generated with the Monte Carlo program of
Zeppenfeld~[1,12], in which the processes were generated to leading order, and
higher order QCD effects on the cross section were approximated by a factor
$K=1+\frac{8}{9}\pi\alpha_s=1.34$. We included the $p_T$ distribution of the
{\sc isajet} $WW$ events in the simulation of the $WW(WZ)$ production. We
calculated the total efficiency for SM couplings to be $0.15\pm 0.02$ for $WW$
and $0.16\pm 0.02$ for $WZ$. The error is 13\%, which is the addition in
quadrature of the uncertainties on the electron trigger and selection
efficiencies (2\%), on the $W(Z)\to jj$ efficiency due to the difference
between the {\sc isajet} and {\sc pythia} programs (9\%), the statistics of the
Monte Carlo samples (4\%), $\rlap{\,/}E_T$ smearing (6\%) and jet energy scale
(6\%). Therefore, $3.2\pm 0.6$~$WW$ and $WZ$ events are expected based on the
SM ($2.8\pm 0.6$~$WW$ events plus $0.4\pm 0.1$~$WZ$ events). The error here is
the sum in
\noindent
quadrature of the uncertainty in the efficiency above and that of the
higher order QCD corrections to the signal prediction (14\%)~\cite{CDF:WW}.

The background estimate, summarized in Table~I, includes contributions from:
multijet events, where a jet was misidentified as an electron and there was
significant (mismeasured) missing transverse energy; $W+\geq 2j$ events with
$W\to e\nu$; $t\bar{t}\to W^+W^-b\bar{b}\to e\nu jjX$; $WW(WZ)$ with $W\to
\tau\nu$ followed by $\tau\to e\nu\bar{\nu}$; and $ZX\to eeX$ where one
electron was not identified.

\begin{figure}[htbp]
\centerline{\psfig{file=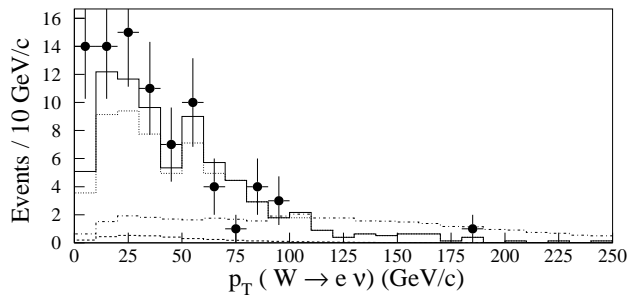
,bbllx=0pt,bblly=0pt,bburx=750pt,bbury=400pt,width=8.5cm}}
\caption{$p_T$ distributions of the $e\nu$ system: data (points), total
background (solid), $W+\geq 2$ jets background (dotted), and Monte Carlo
predictions of $WW$ and $WZ$ production with SM (dashed) and non-SM
(dot--dashed, $\Delta\kappa_Z=\Delta\kappa_\gamma=2$,
$\lambda_Z=\lambda_\gamma=1.5$) couplings.}
\label{PTW_DATA}
\end{figure}

The multijet background was estimated from the data by measuring the
$\rlap{\,/}E_T$ distribution of a background-dominated sample, which was
obtained by selecting events containing an EM cluster that failed at least one
of the electron identification requirements (ii) to (iv) described previously
(shower shape, isolation or track-match). The $\rlap{\,/}E_T$ distribution of
this sample was scaled to match the candidate sample in the region
0~GeV~$<\rlap{\,/}E_T<15$~GeV (before the $\rlap{\,/}E_T$ requirement was
applied), where the contribution of signal events is negligible; then the
portion of this distribution passing the $\rlap{\,/}E_T$ requirement
($\rlap{\,/}E_T>25$~GeV) was taken as our estimate of the multijet background,
giving $12.2\pm 2.3(\rm{stat.})\pm 1.1(\rm{syst.})$ events. The $W+\geq 2j$
background was estimated using the {\sc vecbos} Monte Carlo
program~\cite{vecbos}, with $Q^2=m_W^2$, followed by parton fragmentation using
the {\sc isajet} program and a detailed detector simulation. We normalized the
number of {\sc vecbos} $W+\geq 2j$ events to the number of observed $W+\geq 2j$
events (after multijet events were subtracted) outside of the dijet mass signal
region. This yielded $62.2\pm 8.2(\rm{stat.}) \pm 10.1(\rm{syst.})$~$W+\geq 2j$
background events inside the $m_{jj}$ signal region, where the statistical
uncertainty is due to the size of the {\sc vecbos} $W +\geq 2j$ event sample
and of all samples used to calculate the normalization factor (13\%), and the
systematic error is due to the normalization and to the jet energy scale
correction (16\%). The $W+\geq 2j$ cross section obtained with this
normalization procedure was consistent with the {\sc vecbos} prediction. The
backgrounds due to $t{\bar t}\to W^+W^-b{\bar b}$~\cite{laenen}, $WW(WZ)\to
\tau\nu jj$~\cite{Hagiwara} and $ZX\to eeX$ were estimated using the {\sc
isajet} program followed by the {\sc geant} detector simulation and found to be
small. The total background from all sources was estimated to be $75.5\pm 13.3$
events. Therefore, we observed no statistically significant signal above the
background. Figure~\ref{PTW_DATA} shows the $p_T$ distribution of the $e\nu$
system of the data events, background estimates, and Monte Carlo predictions of
$WW$ and $WZ$ production for SM couplings and for one example with anomalous
couplings.

Using the efficiencies for SM $WW$ and $WZ$ production, the
background-subtracted signal, the branching fractions $B(W\rightarrow e\nu)$,
$B(W\rightarrow$~hadrons) and $B(Z\rightarrow$~hadrons) from Ref.\cite{pdg},
and assuming the SM ratio of the cross sections of $p\bar{p}\to W^+W^-X$ and
$p\bar{p}\to W^\pm ZX$, we set an upper limit at the 95\% confidence level (CL)
on the cross section $\sigma(p\bar{p}\to W^+W^-X)$ of 183~pb.


The absence of an excess of events with high $p_T^W\!$ excludes large
deviations from the SM couplings. To set limits on the anomalous coupling
parameters, we performed a binned likelihood fit on the entire $p_T^W\!$
spectrum. For each $p_T^W\!$ bin in the likelihood fit, and for a given set of
anomalous coupling parameter values, we calculated the probability for the sum
of the background and the predicted signal to fluctuate to the observed number
of events. The uncertainties in the efficiency, background estimates,
integrated luminosity and higher order QCD corrections to the signal prediction
were convoluted in the likelihood function with Gaussian distributions. Our
analysis is sensitive to anomalous couplings at both large and small $\hat{s}$,
since we do not require high $p_T^W\!$, and therefore we retain events with
small $\hat{s}$~\cite{ucla}.

\begin{figure}[htbp]
\centerline{\hbox{
\psfig{bbllx=00pt,bblly=110pt,bburx=545pt,bbury=625pt,figure=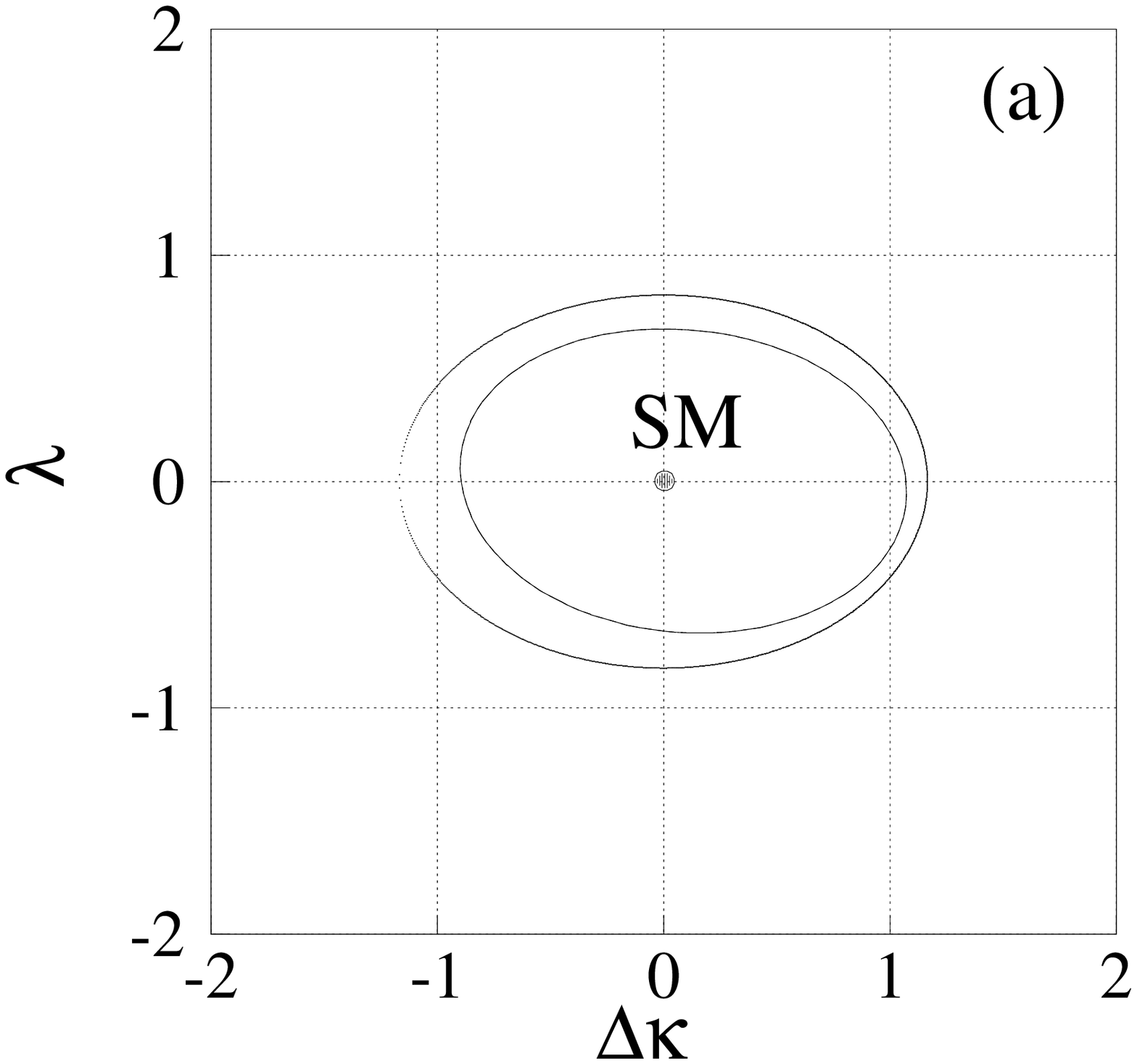,width=1.8in,height=1.7in}
\psfig{bbllx=35pt,bblly=110pt,bburx=580pt,bbury=625pt,figure=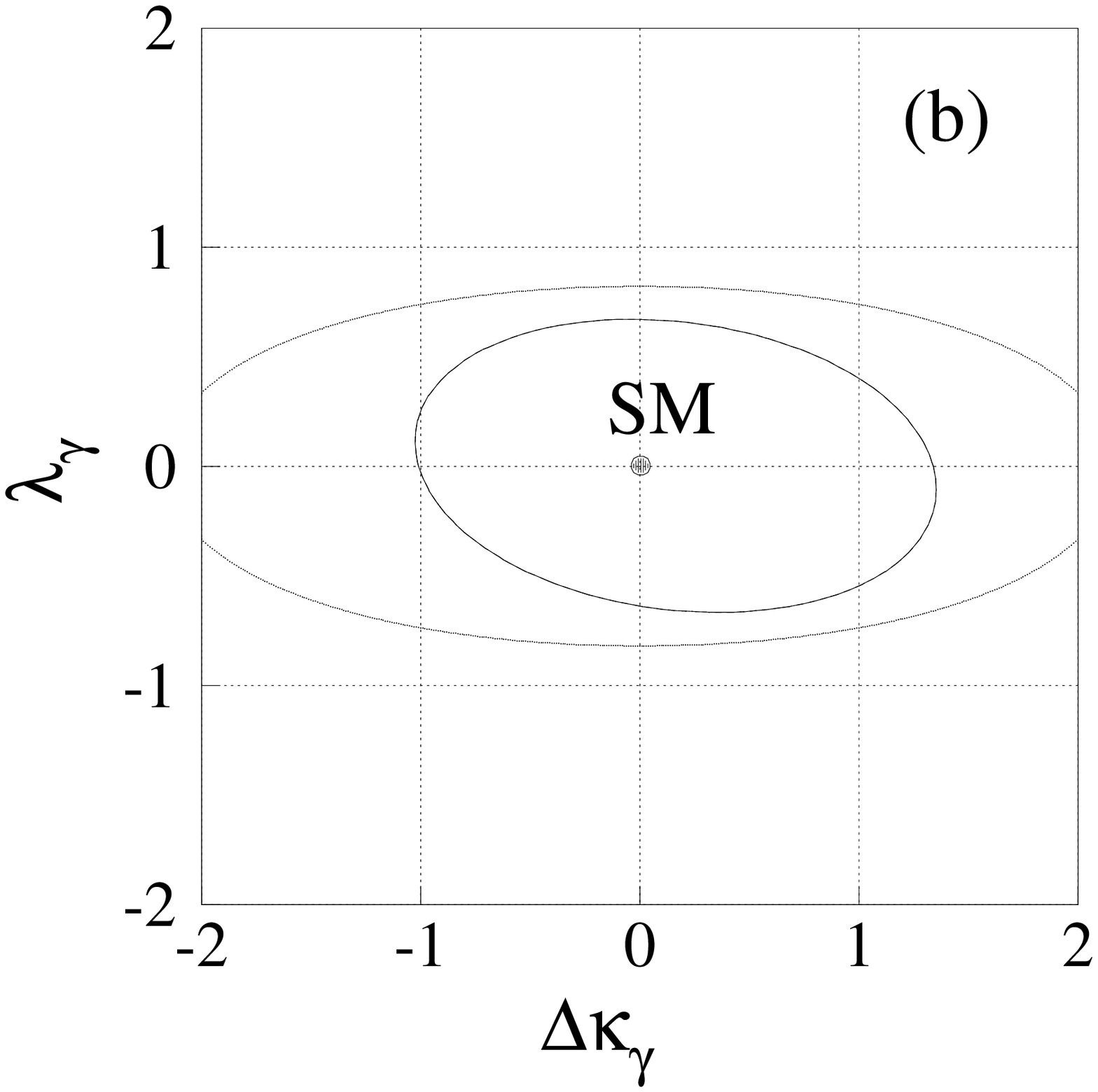,width=1.8in,height=1.7in}}}
\centerline{\hbox{
\psfig{bbllx=00pt,bblly=110pt,bburx=545pt,bbury=625pt,figure=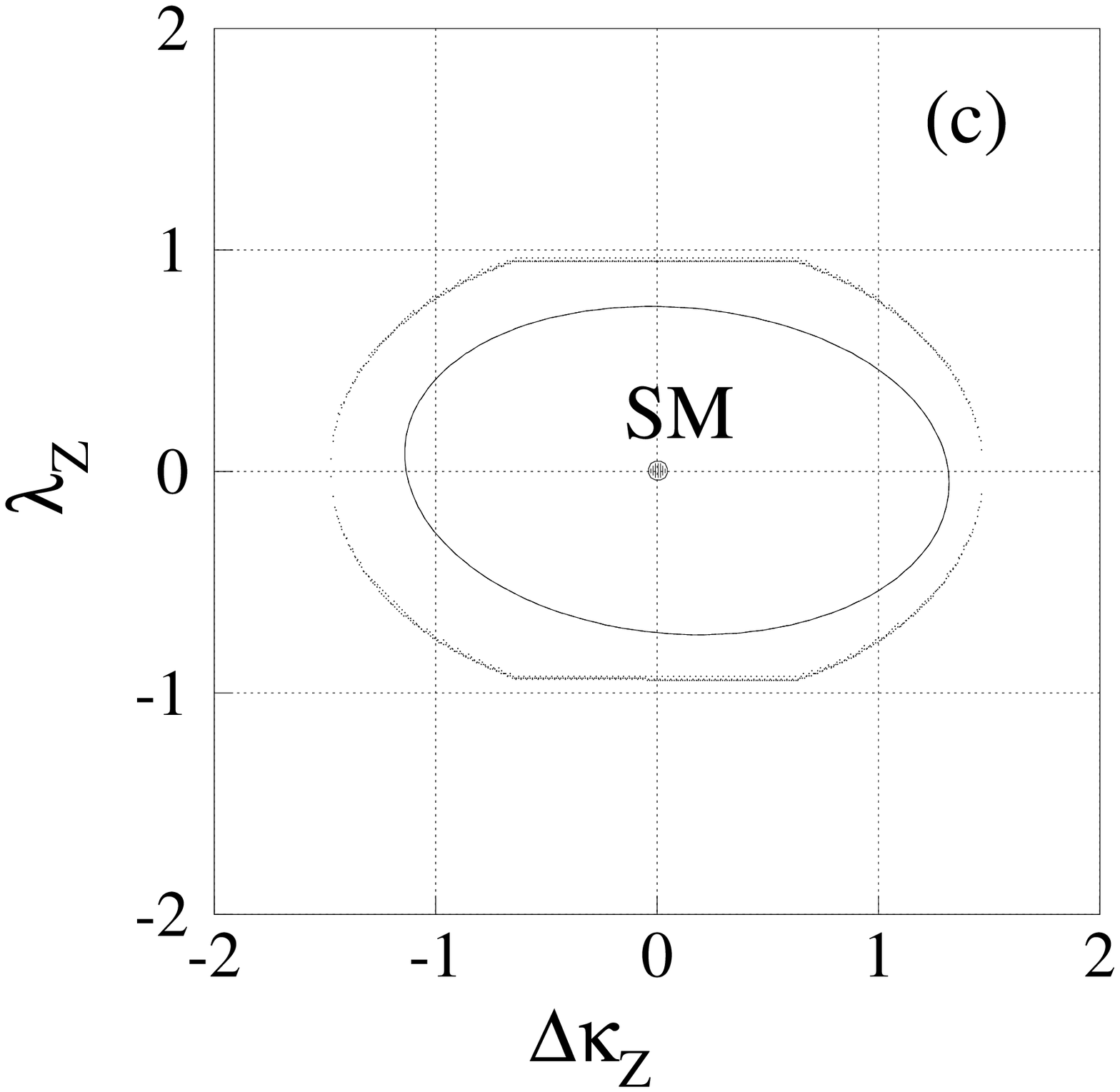,width=1.8in,height=1.7in}
\psfig{bbllx=35pt,bblly=110pt,bburx=580pt,bbury=625pt,figure=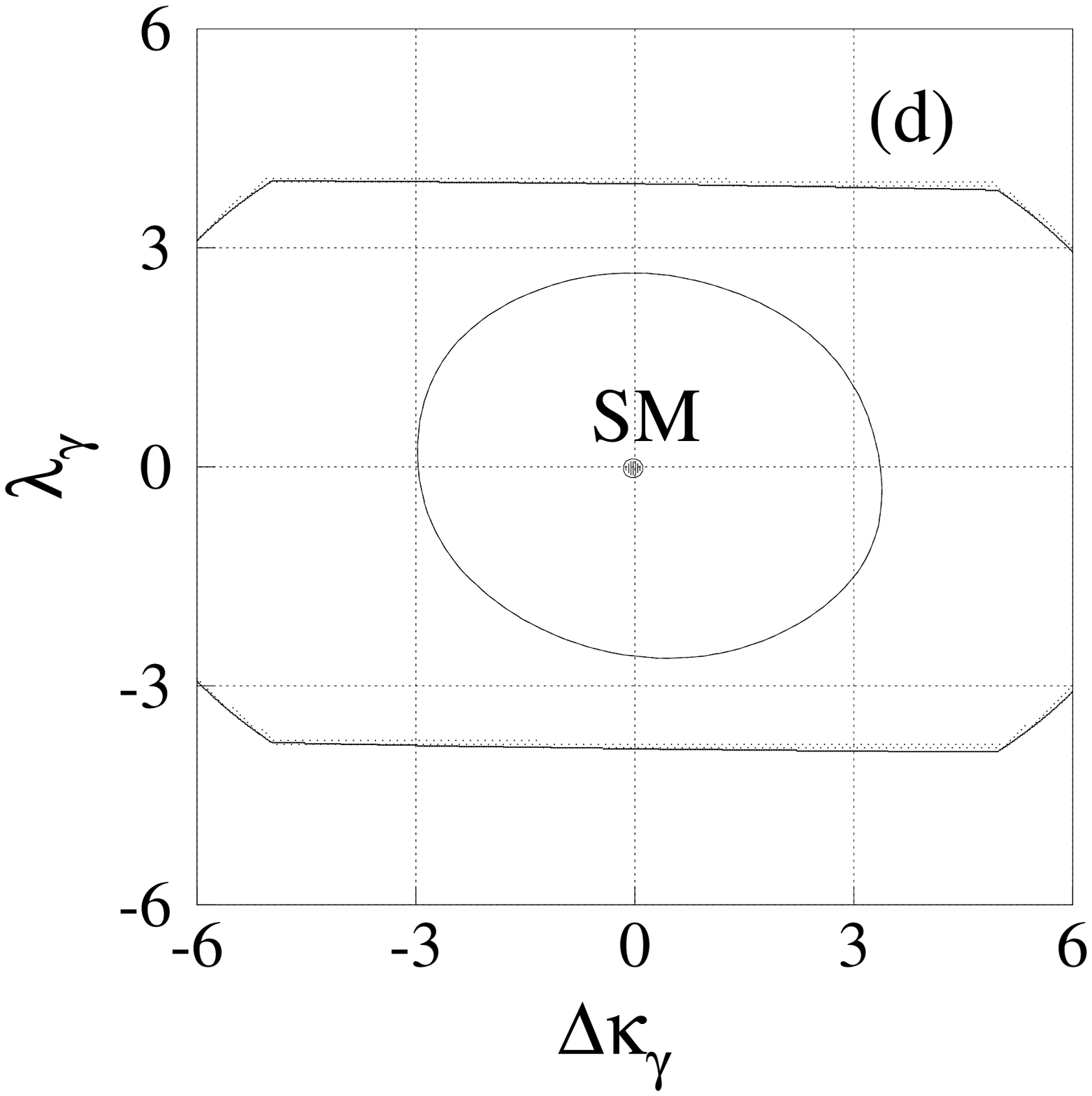,width=1.8in,height=1.7in}}}
\caption{Contour limits on anomalous coupling parameters at the 95\% CL (inner
curves) and limits from S--matrix unitarity (outer curves), assuming
(a)~$\Delta\kappa\equiv\Delta\kappa_{\gamma}=\Delta\kappa_{Z}$,
$\lambda\equiv\lambda_{\gamma}=\lambda_{Z}$; (b)~HISZ relations [17]; (c)~SM
$WW\gamma$ couplings, and (d)~SM $WWZ$ couplings. $\Lambda=1.5$~TeV is used for
(a), (b) and (c); $\Lambda=1.0$~TeV is used for (d). The SM prediction is
$\Delta\kappa=0, \lambda=0$.}
\label{WW:contour}
\end{figure}

We obtained limits on the coupling parameters using four different assumptions
of how the parameters are related to each other (a--d, below). The 95\% CL
contour limits are shown as the inner curves in Fig.~\ref{WW:contour}, along
with the S--matrix unitarity limits, shown as the outer curves, which are
obtained by evaluating all (i.e., $W\gamma$, $WW$ and $WZ$) processes. The 95\%
CL limits on the axes are listed in Table~II. Two $\Lambda$ values are used,
$\Lambda=1.5$~TeV for (a), (b) and (c), and $\Lambda=1.0$~TeV for (d). These
are values used in other measurements [3,5]. For (d), the results obtained with
$\Lambda=1.5$~TeV violate the S--matrix unitarity limit (not shown).

In conclusion, we have searched for anomalous $WW$ and $WZ$ production in the
$e\nu jjX$ decay mode, and set limits on the $WW\gamma$ and $WWZ$ anomalous
coupling parameters $\Delta\kappa$ and $\lambda$. The limits on $\lambda$ are
comparable to those measured using $W\gamma$ production, whereas those on
$\Delta\kappa$ from this analysis are significantly better. The results with
assumption (c) are unique to $WW(WZ)$ production since the $WWZ$ couplings are
not accessible with $W\gamma$ production. The limits on both $\Delta\kappa$ and
$\lambda$ are significantly tighter than those from the analysis using $WW$ to
dilepton decay, due to the additional $WZ$ production mode here, and to the
larger branching fractions of $W(Z)$ decaying to hadrons than that of $W$ to
$e\nu$ or $\mu\nu$ in the dilepton analysis.


\vspace{0.2in}

We thank U. Baur for providing us with much helpful advice and D. Zeppenfeld
for the $WW$ and $WZ$ Monte Carlo generators and useful instructions.  We thank
the staffs at Fermilab and the collaborating institutions for their
contributions to the success of this work, and acknowledge support from the 
Department of Energy and National Science Foundation (USA), Commissariat \` a
L'Energie Atomique (France), Ministries for Atomic Energy and Science and
Technology Policy (Russia), CNPq (Brazil), Departments of Atomic Energy and
Science and Education (India), Colciencias (Colombia), CONACyT (Mexico),
Ministry of Education and KOSEF (Korea), CONICET and UBACyT (Argentina), and
the A.P. Sloan Foundation.

\bibliographystyle{unsrt}


\begin{table*}[h]
\caption{Summary of $e\nu jj$ data and backgrounds.}
\begin{tabular}{llcr}
 & Background source     & $e\nu jj$ events & \\
 & $~~~{\rm multijets}$  & $12.2\pm 2.6$    & \\
 & $~~~W+\geq 2j$        & $62.2\pm 13.0$   & \\
 & $~~~t\bar{t}(m_t=180{\ \rm GeV}/c^2)$   & $0.87\pm 0.12$  & \\
 & $~~~WW, WZ\to \tau\nu jj$ & $0.22\pm 0.02$    & \\
 & $~~~ZX\to eeX$        & $0.00^{+0.34}_{-0.00}$ & \\
 & Total Background      & $75.5\pm 13.3$   & \\
\hline
 & Data                  & 84               & \\
\hline
 & SM $WW+WZ$ prediction & $3.2\pm 0.6$     & \\
\end{tabular}
\end{table*}

\vspace{0.5in}

\begin{table*}[h]
\caption{95\% CL limits with various assumptions.}
\begin{tabular}{lcc}
  Assumptions & $\Lambda$ (TeV) & Limits on axes \\
\hline
  (a) $\Delta\kappa_{\gamma}=\Delta\kappa_{Z}$   & 1.5 &
      $-0.9<\Delta\kappa<1.1~~(\lambda=0)$ \\
\hspace{0.70cm}
      $\lambda_{\gamma}=\lambda_{Z}$
      & & $-0.6<\lambda<0.7~~(\Delta\kappa=0)$ \\
\hline
  (b) HISZ~\cite{hisz}                           & 1.5 &
      $-1.0<\Delta\kappa_\gamma<1.3~~(\lambda_\gamma=0)$ \\
      & & $-0.6<\lambda_\gamma<0.7~~(\Delta\kappa_\gamma=0)$ \\
\hline
  (c) SM $WW\gamma$                              & 1.5 &
      $-1.1<\Delta\kappa_Z<1.3~~(\lambda_Z=0)$ \\
      & & $-0.7<\lambda_Z<0.7~~(\Delta\kappa_Z=0)$ \\
\hline
  (d) SM $WWZ$                                   & 1.0 &
      $-2.8<\Delta\kappa_\gamma<3.3~~(\lambda_\gamma=0)$ \\
      & & $-2.5<\lambda_\gamma<2.6~~(\Delta\kappa_\gamma=0)$ \\
\end{tabular}
\end{table*}


\vspace{0.5in}

\newpage

\end{document}